\documentclass[12pt,a4paper]{article}
\usepackage[english]{babel}
\usepackage{epsfig,graphicx}
\usepackage{amsmath,amssymb}
\usepackage{supertabular,longtable}
 
\usepackage{float}
\makeatletter
\let\table\@btab
\makeatother
 
\begin{document}

\title{Event Indexing Systems for Efficient Selection and Analysis of HERA Data} 

\author{L.A.T. Bauerdick\footnote{
    now at {\em Fermi National Accelerator Laboratory, Batavia, USA}},
        Adrian Fox-Murphy,
        Tobias Haas\footnote{
    on leave at {\em Stanford Linear Accelerator Laboratory, Stanford, USA}}, \\
        Stefan Stonjek and 
	Enrico Tassi\footnote{
    now at {\em NIKHEF, Amsterdam, The Netherlands}}, \\
        {\em Deutsches Elektronensynchrotron, Hamburg, Germany}}
\null\vfil
{\let\newpage\relax 
  \maketitle}
\thispagestyle{empty}

 
\maketitle
\thispagestyle{empty}

\begin{abstract}
  The design and implementation of two software systems introduced
to improve the efficiency of offline analysis of event data taken with the
ZEUS Detector at
  the HERA electron-proton collider at DESY are presented. Two
  different approaches were made, one using a set of event directories
and the other using a tag database based on a commercial object-oriented
database management system. 
These are described and compared. Both
systems provide quick direct access to individual collision events in a
sequential data store of several terabytes, and they both considerably
improve the event analysis efficiency. In particular the tag database
  provides a very flexible selection mechanism and can dramatically reduce 
the computing time needed to extract small subsamples from
  the total event sample. Gains as large as a factor 20 have been obtained.
\end{abstract}
\vfil\vfil\vfil\null
 
\clearpage


\section{Introduction}
\label{sec:intro}

Large High Energy Physics (HEP) detectors typically have many hundreds of
thousands of readout channels and record very large data
samples. The task of storing and managing these data is a challenge
which requires sophisticated data management techniques. Initially the data
provided by the on-line data aquisition system of the detector
must be recorded at typical rates of several megabytes per second.
Subsequently fast access to the entirety of the data must be provided
for reconstruction and analysis.

Various techniques have been employed by different experiments to meet
these requirements. Typically the data are stored in sequential format on
magnetic tapes inside a robotic tape storage and access system containing
thousands of tape cartridges.  The tapes are mounted on a tape 
drive automatically without human intervention, both for reading and writing
the data. Currently, typical tape robots provide storage space for up to
several hundred terabytes of data. The data accumulated
by the ZEUS experiment at the HERA $ep$ collider~\cite{ZEUS,HERA}
over eight years of operation amount to approximately 35
terabytes. In addition, approximately 40 terabytes of simulation data
have been accumulated.

Tape storage systems of the type described above work efficiently when a
large fraction of the data to be retrieved is stored on a single tape. This
is typically the case for targeted simulation data, but is generally not
the case for real event sets when the subset of events required for a 
particular analysis may be very sparse.
When only a small fraction of the events are required and these are spread
out over a number of entire tapes, these systems become inefficient. The
inefficiency originates
both from access to the tapes, typically limited by mechanical
constraints
in the tape robotics systems, and from access to data on individual
tapes, limited by the sequential nature of the data format.  The
sequential format requires large amounts of data to be read from the
tape into the memory of an analyzing computer system in order to
extract the desired information.

Various approaches have been used to address this problem. A standard
solution involves splitting the data at an early stage into many data
samples, often overlapping, according to the foreseen needs of
different physics analyses. The split samples are then stored on
magnetic tapes or on disks. In either case the data can be
analysed efficiently if a high proportion of the events stored in a
given sample are required for a particular analysis. However, this method
has two disadvantages. Firstly,
the data samples from selections for different physics interests will be
overlapping, requiring more total storage
space than the original sample. Secondly, the criteria used to split
the data must be defined at an early stage when the understanding
of the data may still be rudimentary. As a result, the splitting may
have to be repeated several times as the understanding of the data
advances.

The limitations of this method can be avoided if the data are stored
using a database management system with appropriate indexing and query
facilities. However, conventional database management systems such as the
relational database ORACLE\cite{ORACLE} have not yet been able to cope
with the typical data recording and analysis requirements of large HEP
experiments. In particular, in these systems the time needed to
retrieve a single event from the global event
sample may exceed the computing time needed to analyse the event by
orders of magnitude.

In this paper we describe a system which overcomes the limitations
described above. The system is built on top of a standard datastore
consisting of sequential datafiles stored on magnetic disks or tapes,
and uses a commercial object-oriented database management system to
provide the missing index and query facilities.  The system was
designed and implemented for the ZEUS experiment but it could be
adapted for use at other large high energy physics experiments
in operation or under construction.


\section{The Data Recording and Analysis Environment of the ZEUS Experiment}
\label{sec:datareq}

ZEUS is a general-purpose experiment at DESY studying
electron-proton collisions at high energies in the HERA
electron-proton collider. The experiment was designed and built and is
being operated by an international collaboration of 50 institutions
and more than 400 physicists. The experimental program is broad and
ranges from studies on parton dynamics in the nucleon to searches for
new exotic phenomena.

The ZEUS experiment has approximately 300000 electronic channels and
operates with the beam crossing time of the HERA collider of $96~{\rm
  ns}$. For every inverse nanobarn ($1 {\rm nb^{-1}}$) of integrated
luminosity delivered, ZEUS records of the order of 1000 $ep$ collision
events.  At the design luminosity of HERA of $1.5\times 10^{-2}~{\rm
  nb^{-1}s^{-1}}$ this corresponds to a data rate of 15 collision
events per second. Between 1992 and 1999, 130 million events were
collected. With a data size of approximately 100 kilobytes per event before
reconstruction, 13 terabytes of raw data have been written during that
period.

\begin{figure}[htbp]
  \begin{center}
    \epsfig{file=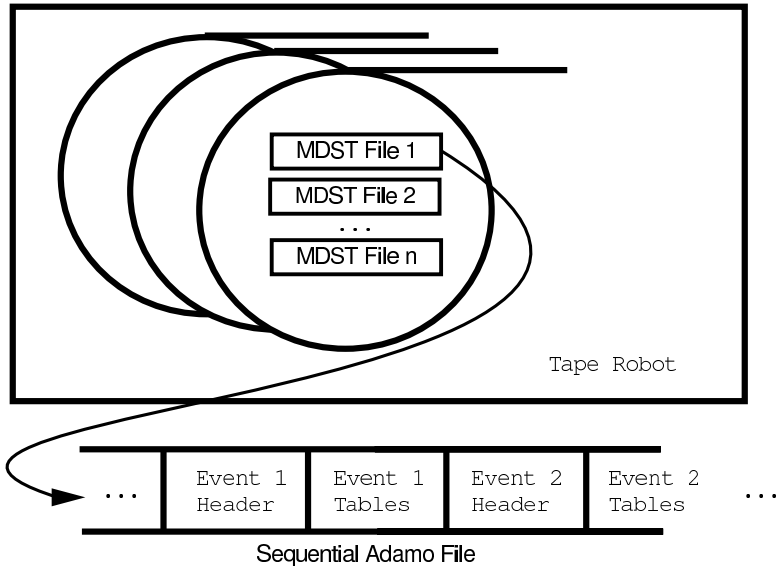,height=7.5cm}
    \caption{Storage model for sequential ZEUS data on magnetic tapes in ADAMO format.}
    \label{fig:adamofiles}
  \end{center}
\end{figure}

Figure~\ref{fig:adamofiles} shows a schematic diagram of the storage model
for
ZEUS data. The data are stored as sequential files on magnetic tape
cartridges in a tape robotic system. The internal data format is
defined by the ADAMO data system~\cite{ADAMO}. ADAMO uses an
Entity-Relationship data model with simple indexing and does not have
query facilities.

The data are reconstructed within a few days after they are recorded.
The delay is required in order to generate calibration constants for the
reconstruction program. The reconstruction program writes two sets of
sequential
ADAMO files. The first set, known as RDST (Reconstructed Data Summary
Tape),
contains the complete raw data information and the result from
reconstruction.  The second set, known as MDST (Mini Data Summary Tape),
is a version of the RDST which is optimized for physics data analysis
where the raw data information is removed. While the data in RDST
format occupy about 150 kilobytes per event, the data in MDST format
are reduced to 25 kilobytes per event. In total, the RDST and MDST
data samples occupy 20 terabytes on magnetic tape cartridges.

Data access for physics analysis is provided through the ZEUS tape
file system (TPFS)~\cite{TPFS,SHIFT}. TPFS defines a
location-independent name space for all data files. When a named data
set is requested by an analysis program, TPFS looks for the
corresponding file in a data storage pool consisting of magnetic disks.  If
the file is not found, TPFS copies the file from the tape store to the
disk pool and makes it available for reading by the requesting
analysis program.  TPFS removes files from the disk pool which have
not been used in the last few days, while files which are required
frequently are kept permanently on disk. In particular, all MDST data
are permanently available. For the data taken from 1992 through 1999
the disk pool size is 3 terabytes.

\begin{figure}[htbp]
  \begin{center}
    \epsfig{file=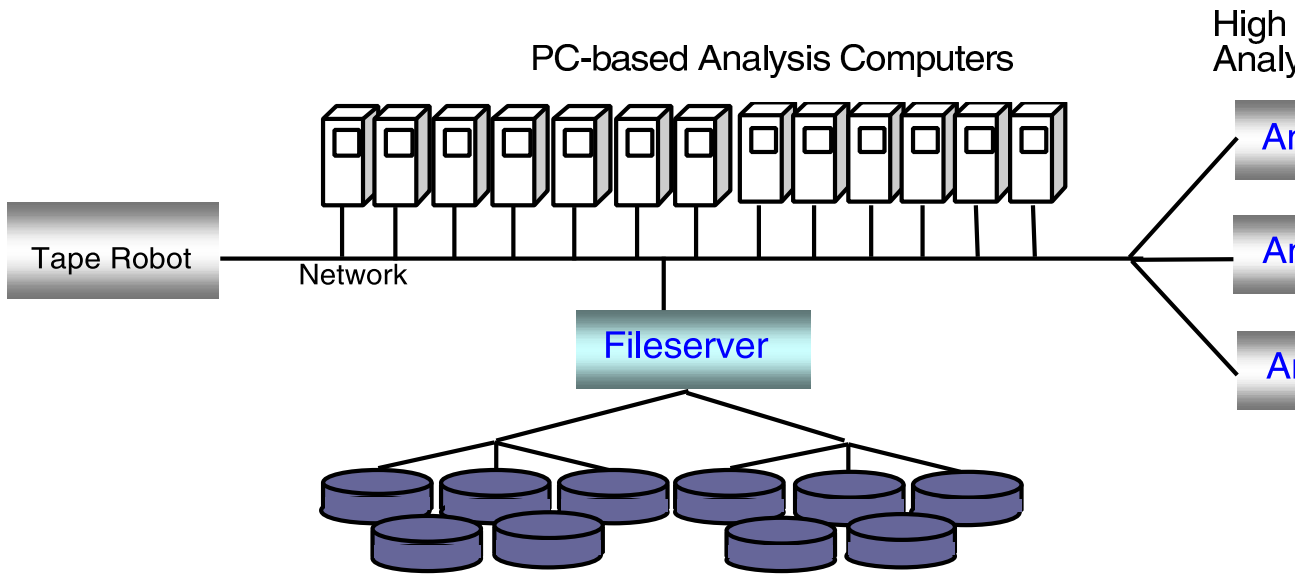,height=7.5cm}
    \caption{Configuration of the analysis computer system of the ZEUS
      Experiment.}
    \label{fig:zarahconf}
  \end{center}
\end{figure}

Figure~\ref{fig:zarahconf} shows the configuration of the computer
system used for analysis of the ZEUS data. The magnetic disks are
connected to a central computer that acts as a fileserver. Many
different computer systems used for running analysis programs
communicate with the fileserver via high-throughput network
connections (100MBit Ethernet, GigaBit Ethernet and HIPPI). The tape
robot system is connected using similar network connections.


\section{Event Directories}
\label{sec:evdir}

In order to remedy the limitations of sequential access to the event
data, a system was developed for ZEUS in which single events in the
event store are accessed directly using a system of {\it event
directories}. An event directory is an index containing for each event
128 logical event flags as well as the collider run number\footnote{A
run is a datataking period with identical trigger conditions. One run
typically lasts a few hours and contains up to 300,000 events.}, the
event number, and the location of the event in the sequential event
store.  The event flags are determined once during reconstruction and
indicate whether the topology of the event matches each one of a wide
range of physics conditions.

\begin{figure}[htbp]
  \begin{center}
    \epsfig{file=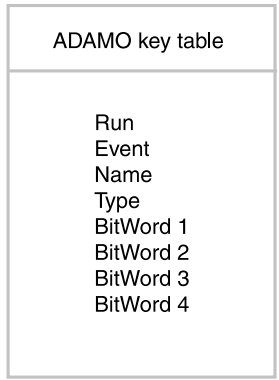,height=5cm}
    \caption{ADAMO key table. A copy of the indicated quantities are stored for
      each record (i. e. event) in the database together with the
      offset in the sequential event file.}
    \label{fig:zdskey}
  \end{center}
\end{figure}

The event directory system uses the capability of the ADAMO system to
index records in a sequential data file. This index is implemented
using a key table. Figure~\ref{fig:zdskey} shows the structure of the
ADAMO key table. In addition to name and type of record, and run and event
number, the table permits storage of four more 32-bit quantities. These
are used in the event directory system to encode the 128 event flags.

\begin{figure}[htbp]
  \begin{center}
    \epsfig{file=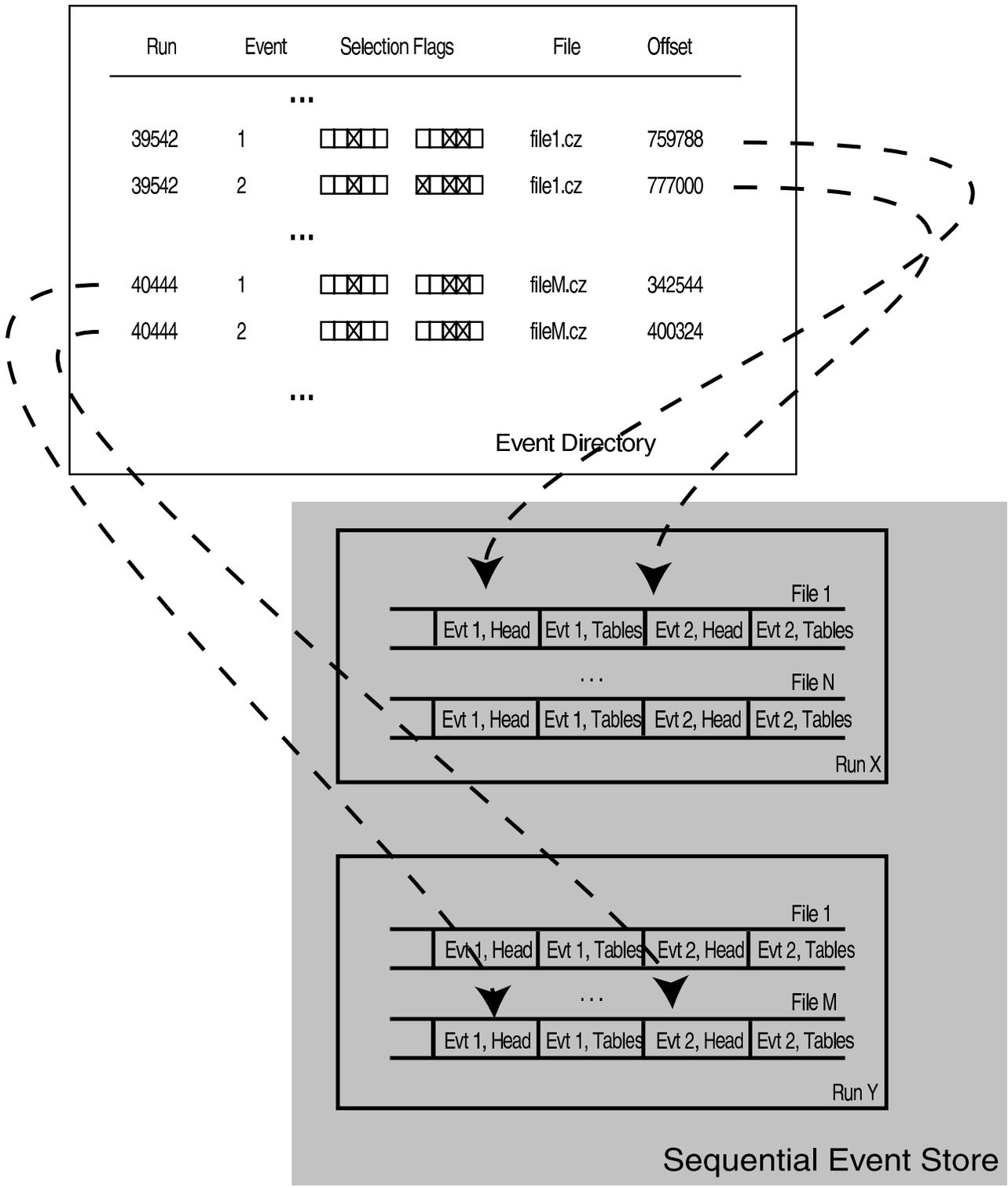,height=20cm}
    \caption{Schematic of the ZEUS event directory system.}
    \label{fig:evdir}
  \end{center}
\end{figure}

Figure~\ref{fig:evdir} depicts schematically how event directories are
used to access events. The event directory information is stored
run-by-run and event-by-event in tables. The tables can be queried
from an analysis job. For example a user might request all events where
certain conditions on the event flags are fulfilled. The event
directory system searches through the event directory tables and for
every requested entry it locates the event file, positions the file
pointer at the proper offset and reads and decodes the event record.
Then control is given to the analysis code to perform whatever
data analysis the user wants to perform.
For convenience, event directory files are stored in a human readable
format. An excerpt from an event directory file is given in
Appendix~\ref{app:b}.

The event directory system was developed for the ZEUS experiment and
has been in operation since 1994, working reliably and efficiently.
The CPU time overhead required to read ZEUS events through the event
directory system is small compared to the event reading time, and much
smaller than with the sequential method of reading events.


\section{The Tag Database}
\label{sec:tagdb}

While event directories work very efficiently, the event selection power is
limited to boolean combinations of precalculated event flags. Furthermore,
it is only possible to test those selection conditions which were
considered when the event flags were calculated.  Also, if quantities on
which the selection condition were based have changed -- e. g. due to
recalibration -- the selection may be invalid. This happens quite
frequently since analysis methods and detector understanding are evolving
quickly. The event flags must be recalculated after every change in the
selection procedure. For large data samples this amounts to a major effort
that can be afforded only a few times per year.

The number of recalculations can be reduced if more information is
stored with each event in the event directory. This information would then
be updated when for instance calibration has changed. As more information
is stored, the event selection becomes more flexible. For instance, rather
than setting a flag when a vertex is found, the reconstructed position of
the vertex can be stored instead. In this case, with appropriate database
technology, one can select not only those events that have a vertex but
also those with a vertex within certain bounds.

Such a system is known as a {\it tag database} in order to stress its
database character and its indexing capabilities and thus distinguish it
from the simpler event directories.
 
The storage requirement for a tag database which stores 200 32-bit
quantities for each event in a data sample of 100 million events is 80
gigabytes. In order to be efficient such a system requires advanced
database management technology. In particular, the CPU time overhead to
retrieve one event from the system must be kept small compared to the time
needed to read the sequential event information.

The need for a tag database within the ZEUS experiment became apparent in
1996 when it was recognized that data analysis would have to become
much more efficient in order to cope with the ever growing data
samples of the experiment. A project was initiated
to design and build a tag database with the following goals:
\begin{itemize}
\item Provide at least the functionality of the
event directory system with equal or greater efficiency.
\item Substantially improve the selection capabilities
  compared to those available with the event directory system.
\item Allow for growth. The database technology should not
  limit future growth of the system. In particular, the system was
required to be capable of handling event samples of several terabytes in
size and to be capable of storing not only the tag information but also the
entire data of the experiment, which may be required in the future.
\item Provide an implementation backwards-compatible with the
  existing system and require only minimal changes to the physics analysis
  codes.
\item Ensure that in addition to serving as an event index, the tag
database be usable standalone as a compact data sample.
\item Allow simple maintenance of the system. In particular, it was
required to be able to partially update the database quickly when
needed.
\end{itemize}

\section{Implementation of the ZEUS Tag Database}
\label{sec:impl}

The ZEUS tag database was implemented using the Objectivity/DB database
management system. This commercial software product of Objectivity
Inc.~\cite{OBJECTIVITY} is an object-oriented database management
system based on the concept of database federations. 
Objectivity/DB can handle databases up to a limit of 10000 petabytes.
 
A number of other HEP experiments are currently using or planning to use
Objectivity/DB for their event storage. An example is the BaBar
experiment at SLAC~\cite{BABAR} which was the first HEP experiment to
choose this product. Other examples are the future experiments at the
CERN LHC, for which the RD45 project at CERN~\cite{RD45}
had studied databases with sizes of up to several terabytes in order to
establish that Objectivity/DB could be used for storing the event data
of the LHC experiments. These experiments expect to record data samples of
several petabytes, about two orders of magnitude larger than the
volume expected for the ZEUS event data.

The code for the ZEUS tag database system is written in C++. This was
the natural choice given the selection of an object-oriented database
management system. Since most ZEUS code is written in FORTRAN, it was
necessary to provide an interface layer in order to make the system usable
by all physicists in the experiment with minimal modifications to their
analysis codes. This layer mimics the existing FORTRAN interface layer
between the analysis codes and the data storage.

\begin{figure}[htbp]
  \begin{center}
    \epsfig{file=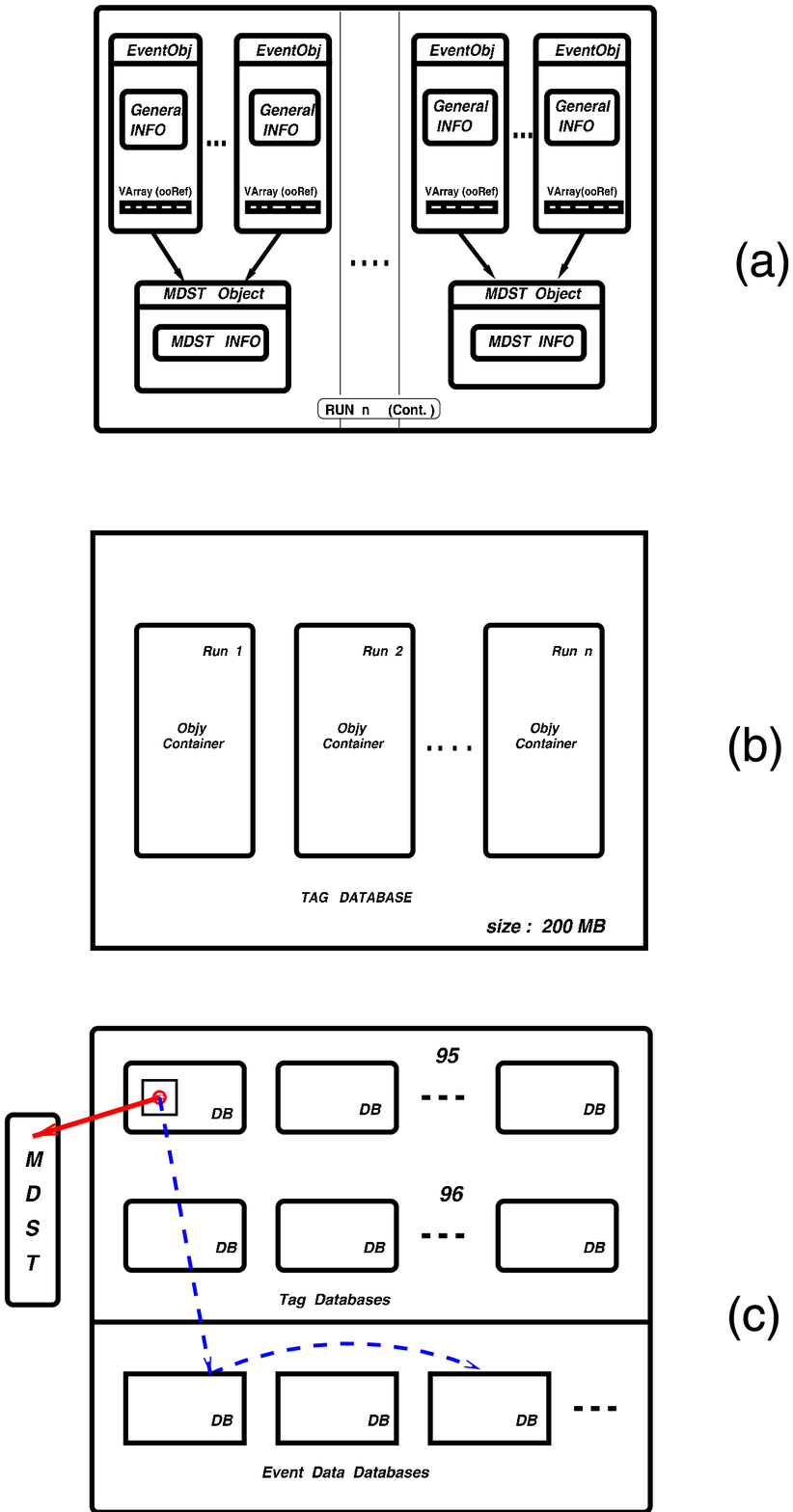,height=23cm}
    \caption{A sketch of the data model of the ZEUS tag database. }
    \label{fig:datamodel}
  \end{center}
\end{figure}

Figure~\ref{fig:datamodel} shows the data model of the tag database at
different levels of detail. Figure~\ref{fig:datamodel}(a) shows event
and MDST objects. An event object contains run and event numbers as
well as a variable-length array of more than 200 event variables with
information on kinematics, identified particles, calorimetry, tracking and
jets, as well as the selection bits used in event directories. An
overview of the stored variables is given in Appendix~\ref{app:a}.
All events of a given run are stored in one {\it container} of
Objectivity/DB. Several of these containers are grouped
together into one {\it database} of the federation. The size of these
databases is kept small ( around 200~MB ) for convenience of data
management. This is illustrated in Figure~\ref{fig:datamodel}(b).
Finally, as shown in Figure~\ref{fig:datamodel}(c), all these
databases make up the {\it federated database}.
Figure~\ref{fig:datamodel}(c) also shows how additional event data
could be stored in the system. This part could be implemented in the
future if needed.


\section{Performance}
\label{sec:perf}

The overhead in computing time generated by an event indexing system
such as either the event directory system or the tag database must be
held small in comparison to the time needed to read and analyse events
from the sequential event store. Both the event directory system and
the tag database were designed with this consideration in mind and their
performance is being
monitored regularly.

Table~\ref{tab:evdirtime} shows measurements of the CPU time overhead
from the event directory system. These times were measured on a
Silicon Graphics Challenge XL computer~\cite{SGI} with R10000
processors running at 194~MHz.
\begin{table*}[htbp]
  \begin{center}
    \begin{tabular}{|c|c|}\hline
      Selection & Time \\
      \hline
      Sequential Read                        & 200 s \\
      Event Directory, No Selection          & 190 s \\
      Event Directory, Selection 1 out of 2  & 190 s \\
      Event Directory, Selection 1 out of 20 & 260 s \\
      \hline
    \end{tabular}
  \end{center}
\caption{Computing time used for reading 10000 events from disk with the ZEUS
  Event Directory system. The times are CPU time measured on a Silicon
  Graphics Challenge XL computer with R10000 processors running at a
  clock rate of 194 MHz. Events were read, but not analysed.}
\label{tab:evdirtime}
\end{table*}
10000 events were read in four different ways. 

In the first measurement (labelled ``Sequential Read'' in
Table~\ref{tab:evdirtime}), all events were read
sequentially from the event store without using the event
directory system. About $20~{\rm ms}$ of CPU time were required to read
one event. In the second case (``Event Directory, No
Selection''), events were accessed using the event directory system
without applying any selection. The time required was slightly smaller
than when the data were accessed without using the event directories.
This is due to the fact that in addition to the event data the sequential
event store contains test and calibration information which in the first
measurement was read and ignored. The event
directory system skips unused non-event data without ever reading it.
In the third measurement (``Event Directory, Selection 1 out of 2''),
a selection was applied which picked approximately one out of every
two events. No significant overhead was observed in this case. Only when a
stronger selection is applied does the event directory overhead become
considerable. This can
be seen in the fourth measurement (``Event Directory, Selection 1
out of 20''), where a selection was applied which selected
approximately one out of every twenty events. The overhead here compared
to the previous measurement was 70 seconds.
Since 200,000 events had to be scanned to select 10,000 events, this
corresponds to a CPU time of $0.3~\rm{ms}$ per scanned event.

\begin{figure}[htbp]
  \begin{center}
    \epsfig{file=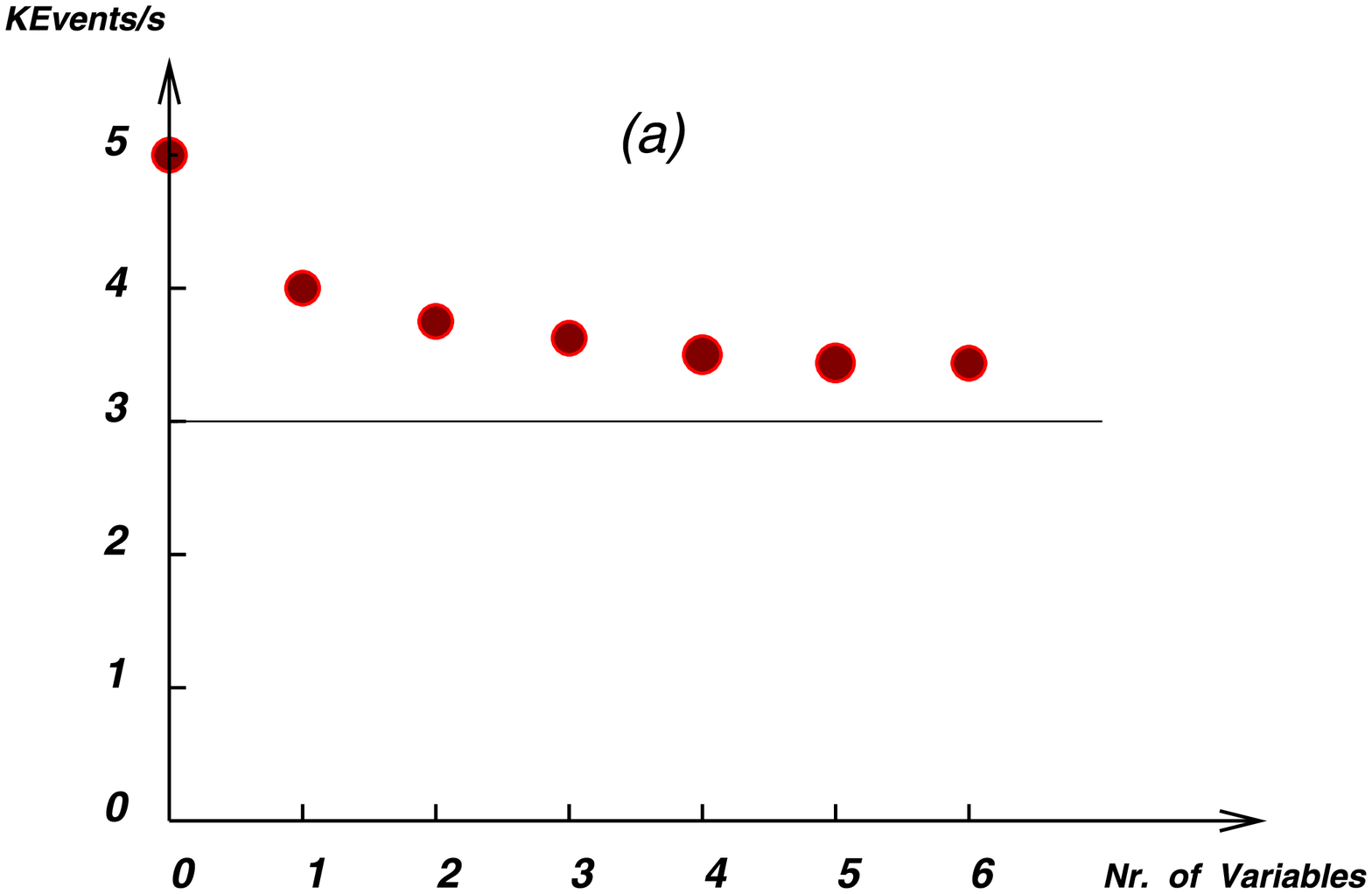,height=11cm,angle=90}
    \epsfig{file=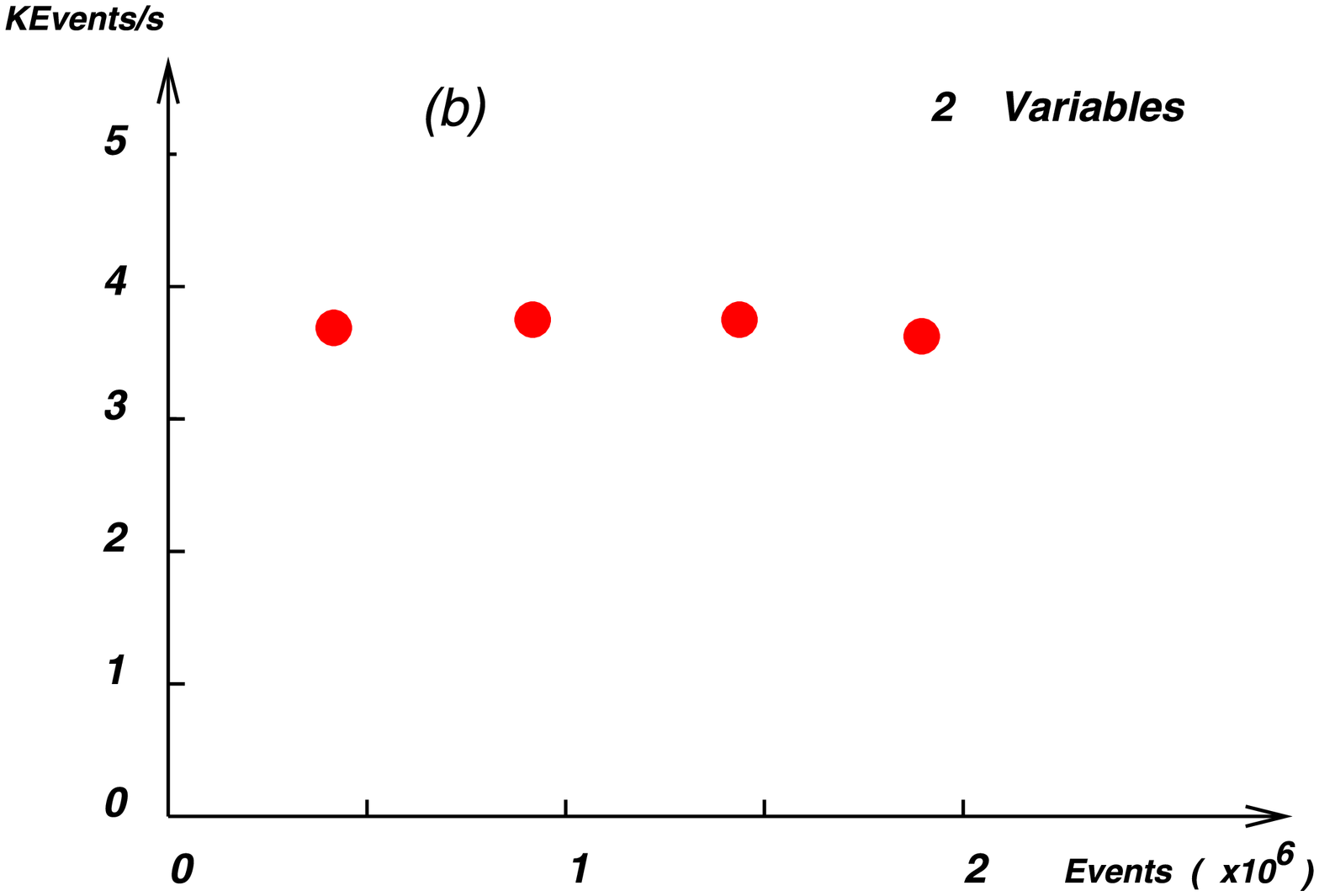,height=11cm,angle=90}
    \caption{The rate of events processed using the ZEUS tag database
      (a) as a function of the number of variables used inside the
      query and (b) as a function of the total number of events stored
      in the tag database. The measurements were done with a prototype
      implementation of the tag database using Objectivity/DB version
      3.8 on a Silicon Graphics Challenge XL machine with R4400
      processors running at a clock rate of 150 MHz.}
    \label{fig:readrate}
  \end{center}
\end{figure}

The performance of the tag database is illustrated in
Figures~\ref{fig:readrate}(a) and (b). 
Figure~\ref{fig:readrate}(a) shows the rate of events processed by the
tag database as a function of the number of variables used in the
query. A maximum number of 6 variables was used for the measurement
since this is a typical number of variables used for data analyses.
For an empty query a rate of 5000 events per second is reached. For
non-empty queries the rate decreases only weakly with increasing the
number of variables in the query, remaining above 3500 events per
second, i.e. $0.28~{\rm ms}$ per event. This shows that the
performance of the tag database is similar to that of the event
directory system, even though much more information is stored in the
database.  Figure~\ref{fig:readrate}(b) shows the rate of events
processed as a function of the total number of events contained in the
tag database.  In this case a query involving two variables was used.
No dependence of the rate on the size of the database is observed.
These two observations confirm that the tag database has a CPU time
overhead which is even smaller than that of the event directory method.
Thus the tag database exceeds the required performance.

The gain in analysis efficiency achieved using either event
directories or the tag database is illustrated in
Table~\ref{tab:perfgain}.
\begin{table*}[htbp]
  \begin{center}
    \begin{tabular}{|c|c|c|c|c|}\hline
      System & Selection & Events scanned & Events selected & Time \\
      \hline
      Sequential Data & No Selection              & 25000 & 25000 &  485 s \\
      \hline
      Event Directory & One electron found        & 25000 & 11000 &  203 s \\
      Tag Database    & One electron found        & 25000 & 11000 &  197 s \\
      \hline
      Event Directory & $E_T > 30~{\rm GeV}$      & 45000 &  2750 &  793 s \\
      Tag Database    & $E_T > 30~{\rm GeV}$      & 45000 &  2750 &  105 s \\
      \hline
    \end{tabular}
  \end{center}
\caption{Comparison of computing time required for different
  selections for event directories and tag database. The times are
  CPU time measured on  a Silicon Graphics Challenge XL computer with 
  R10000 processors running at a clock rate of 194 MHz.Events were 
  read but not analysed.}
\label{tab:perfgain}
\end{table*}
The first row shows the CPU time required to read 25,000 events. About
$20~{\rm ms}$ are needed to read one event from the sequential event
store. The second and third rows show the times required to select the
11,000 events that contain at least one electron candidate from a total
sample of 25,000 events using either event directories or the tag
database. The CPU requirement is almost the same in both cases. It
corresponds again to about $20~{\rm ms}$ per event. The fourth and
fifth rows give the time required to select the 2,750 events with
transverse energy greater than $30~{\rm GeV}$ from a total of 45,000
events. The result from using event directories is shown in the fourth
row. In this case the total time corresponds to the time required to
read all 45,000 events from the sequential event store since the event
directories have no precalculated flags for the query $E_T > 30~{\rm
  GeV}$. The fifth row shows the result when using the tag database.
In this case the tag database is almost an order of magnitude faster.
This is possible since the tag database stores the value of the
transverse energy for each event.  Hence the time used is governed by
the time needed to read in the selected events only. This
illustrates the power of the tag database.


\section{Summary}
\label{sec:summary}

The offline data storage environment of the ZEUS experiment uses a
sequential event store together with two different indexing systems,
an event directory system and a tag database. Both systems were
developed after the experiment had started taking data. They improve
the efficiency of accessing events and hence the efficiency with which
data analyses can be performed.

The event directories and the tag database have different selection
capabilities. Event directories are limited to selections based on
combinations of a set of 128 event flags which are calculated once
during reconstruction of the data. The tag database extends the
selection capabilities substantially. Selection based on the values of
over 200 floating point variables and a large number of flags can be
performed.  Furthermore the tag database can also be used
independently of the sequential event store as a compact standalone
data sample. The tag database could readily be extended to contain more
variables, even to include the complete event information.

The event directories are implemented using the same technology as for
the sequential event store, namely the ADAMO data system. The tag
database uses the commercial object-oriented database management
system ``Objectivity/DB''.

It has been demonstrated that the CPU time overhead introduced by
either system is small -- at the level of 1\% of the total CPU
time required to read complete events. A substantial reduction in
the CPU time required to select events has been achieved.
Using event directories, savings of the order of a factor 2 to 3
have been achieved, while for the tag database savings as large as a
factor 20 are observed.  Thus, the use of the tag database has
dramatically improved the efficiency of data analysis within the ZEUS
collaboration.


\begin{appendix}

\section{List of Physics Analysis Quantities stored inside the ZEUS
  Tag Database }
\label{app:a}

The following list describes the physics quantities that are stored
inside the ZEUS tag database. These quantities can be used to selected
complete events from the event store. They can also be used directly
as a very compact data sample.

\begin{enumerate}
\item Run and Event Number (2 integer variables),

\item Flags:
  \begin{itemize}
  \item First Level Trigger Flags (64 1-Bit quantities),

  \item Second Level Trigger Flags (192 1-Bit quantities),

  \item Third Level Trigger Flags(352 1-Bit quantities),

  \item Offline Event Selection Flags (128 1-Bit quantities; same as Event Directory flags),

  \item Miscellaneous Flags (64 1-Bit quantities),
  \end{itemize}

\item First Candidate for Deep Inelastically Scattered Electron, Algorithm A:
  \begin{itemize}
  \item Calorimeter Energy and Position measurements (13 floating point variables),

  \item Position measurements from other detector components (11 floating point variables),
  \end{itemize}

\item Second Candidate for Deep Inelastically Scattered Electron, Algorithm A (5 floating point variables),

\item First Candidate for Deep Inelastically Scattered Electron, Algorithm B:
  \begin{itemize}
  \item Calorimeter Energy and Position measurements (13 floating point variables),

  \item Position measurements from other detector components (11 floating point variables),
  \end{itemize}

\item Second Candidate for Deep Inelastically Scattered Electron, Algorithm B (5 floating point variables),

\item Estimators for Event Kinematics:
  \begin{itemize}
  \item Using electron from algorithm A (7 floating point variables),

  \item Using electron from algorithm B (7 floating point variables),
  \end{itemize}

\item Global Calorimeter Variables:
  \begin{itemize}
  \item Total energy, transverse energy, missing transverse energy (26 floating point variables),

  \item Energy in different calorimeter parts. (3 floating point variables),

  \item Hadronic four vectors with 2 different methods (8 floating point variables),

  \end{itemize}

\item Tracking Quantities:
  \begin{itemize}
  \item Number of Primary and Secondary Tracks, Vertex Positions (10 floating point variables),

  \item Transverse energy from tracking (5 floating point variables),
  \end{itemize}

\item Luminosity Measurement Information (6 floating point variables),

\item Information from the Muon System (7 floating point variables),

\item Identified Muons (6 floating point variables),

\item Leading Proton Measurement (7 floating point variables),

\item Beampipe Calorimeter Measurement (7 floating point variables),

\item Forward Neutron Measurement (5 floating point variables),

\item Low Angle Tagging Devices (7 floating point variables),

\item Jet Measurement from 4 Different Jet Finders (28 floating point variables),

\item Charmed Mesons, $D^\star$, $D0$, $D_S$ (15 floating point variables).

\end{enumerate}

\section{Example of a ZEUS Event Directory File}
\label{app:b}

\begin{minipage}[t]{14cm}
\tiny
\begin{verbatim}
TABLE 10

[...]

 /* ZEDFILEX (ID, Name(4), Options) */
 1, 'MDST2.D000331.T224552.R035762A.cz', '' , '' , '' , 
    'MEDIUM=COMP,DRIVER=FZ,FILFOR=EXCH,SFGET';
 END TABLE

 TABLE 11
 /* ZEDMETAX (ID, Name, OFF) */
 1, 'HSYOUT'     ,   137;
 2, 'HEAD'       , 62751;
 3, 'MDSTDFL00V0', 63757;
 END TABLE

 TABLE 12
 /* ZEDIRX (ID, GAFTyp, Nr1, Nr2, TStam11, TStam12, TStam21, TStam22, OFF) */
    1, 'EVTF', 35762,   16, X'00000468', X'0000060', X'00000000', X'000000', 62751;
    2, 'EVTF', 35762,   17, X'00000068', X'0000040', X'00000000', X'000000', 90011;
    3, 'EVTF', 35762,   20, X'20000460', X'0002020', X'12000000', X'040000', 102480;
    4, 'EVTF', 35762,   21, X'00000028', X'0000040', X'00000000', X'000000', 131195;
    5, 'EVTF', 35762,   22, X'20008A60', X'0102000', X'00000000', X'000000', 142054;
    6, 'EVTF', 35762,   23, X'00000068', X'0000040', X'00000000', X'000000', 151840;


[...]

}
\end{verbatim}
\end{minipage}
\end{appendix}

\begin{thebibliography}{99} 

\bibitem{ZEUS}
ZEUS Collaboration, M. Derrick et al., {\em A Measurement of sigma/tot
  (Gamma Proton) at sqrt(s)=210 GeV}, Physics Letters B 293 (1992) 465-477.\\ 
ZEUS Collaboration, {\em The ZEUS Detector, Status Report}, DESY
1993.\\
{\tt http://www-zeus.desy.de}.

\bibitem{HERA}
R. D. Peccei (Ed), {\em Proceedings of the HERA Workshop}, DESY 1987.

\bibitem{ORACLE}
{\tt http://www.oracle.com}

\bibitem{ADAMO}
G.Kellner, {\em Development of software for Aleph using structured techniques}, ; Comp. Phys. Comm. 45 (1987).\\
Z.Qian et al., {\em Use of the ADAMO Data Management System within ALEPH}, Comp. Phys. Comm. 45 (1987).\\
H.Kowalski et al., {\em Investigation of ADAMO performance in the ZEUS calorimeter reconstruction program.}, Comp. Phys. Comm. 57 (1989).\\
S.M. Fisher, P. Palazzi, {\em The ADAMO System Programmers Manual}, CERN ECP and RAL. 

\bibitem{TPFS}
O. Manczak, {\em ZARAH Tape File System}, ZEUS Note 96-045, DESY 1996.\\
D. Gilkinson and O. Manczak, {\em The Naming Convention for the ZARAH Data Store}, ZEUS Note 96-046, DESY 1996. 

\bibitem{SHIFT}
A. Buijs, J.D. Hobbs, B. Panzer-Steindel, N.K. Watson and A.M. Lee,\\
{\em  Distributed Physics Analysis in the OPAL Experiment}, Proc. Int. Conf. Comp. High Energy Phys., C. Verker and W. Wojcik, Eds., Annecy 1992.


\bibitem{OBJECTIVITY}
{\tt http://www.objectivity.com}

\bibitem{BABAR}
The BaBar Collaboration, {\em Technical Design Report}, SLAC 1995.

\bibitem{RD45}
J. Shiers, {\em CERN RD45 Status, A Persistent Object Manager for HEP},\\
Proc. Int. Conf. Comp. High Energy Phys., Berlin 1997.

\bibitem{SGI}
{\tt http://www.sgi.com}

\end{thebibliography}
\end{document}